\documentclass[sigconf]{acmart}
\usepackage{multirow}
\usepackage{subcaption}
\usepackage{hyperref}
\AtBeginDocument{%
  \providecommand\BibTeX{{%
    \normalfont B\kern-0.5em{\scshape i\kern-0.25em b}\kern-0.8em\TeX}}}

\setcopyright{acmcopyright}
\copyrightyear{2018}
\acmYear{2018}
\acmDOI{10.1145/1122445.1122456}

\acmConference[Woodstock '18]{Woodstock '18: ACM Symposium on Neural
  Gaze Detection}{June 03--05, 2018}{Woodstock, NY}
\acmBooktitle{Woodstock '18: ACM Symposium on Neural Gaze Detection,
  June 03--05, 2018, Woodstock, NY}
\acmPrice{15.00}
\acmISBN{978-1-4503-XXXX-X/18/06}

\begin{document}

% \title{DLModelCI: A Scalable and Elastic Model Management Platform for DNN Inference Serving in Clouds}
\title{MLModelCI: An Automatic Cloud Platform for Efficient MLaaS}

% \affiliation{%

% \affiliation{%

% \affiliation{%

% \affiliation{%

% \affiliation{%

% \affiliation{%

\author{Huaizheng Zhang}
\affiliation{\institution{Nanyang Technological University}}
\email{huaizhen001@e.ntu.edu.sg}

\author{Yuanming Li}
\affiliation{\institution{Nanyang Technological University}}
\email{yli056@e.ntu.edu.sg}

\author{Yizheng Huang}
\affiliation{\institution{Nanyang Technological University}}
\email{yizheng.huang@ntu.edu.sg}

\author{Yonggang Wen}
\affiliation{\institution{Nanyang Technological University}}
\email{ygwen@ntu.edu.sg}

% \author{Yong Luo}
% \affiliation{\institution{Nanyang Technological University}}
% \email{yluo@ntu.edu.sg}

\author{Jianxiong Yin}
\affiliation{\institution{NVIDIA AI Tech Center}}
\email{jianxiongy@nvidia.com}

% \author{Nguyen Binh Duong TA}
% \affiliation{\institution{Singapore Management University}}
% \email{donta@smu.edu.sg}

\author{Kyle Guan}
\affiliation{\institution{Nokia Bell Labs}}
\email{kyle.guan@nokia.com}


\begin{abstract}

MLModelCI provides multimedia researchers and developers with a one-stop platform for efficient machine learning (ML) services. The system leverages DevOps techniques to optimize, test, and manage models. It also containerizes and deploys these optimized and validated models as cloud services (MLaaS). In its essence, MLModelCI serves as a housekeeper to help users publish models. The models are first automatically converted to optimized formats for production purpose and then profiled under different settings (e.g., batch size and hardware). The profiling information can be used as guidelines for balancing the trade-off between performance and cost of MLaaS. Finally, the system dockerizes the models for ease of deployment to cloud environments. A key feature of MLModelCI is the implementation of a controller, which allows elastic evaluation which only utilizes idle workers while maintaining online service quality. Our system bridges the gap between current ML training and serving systems and thus free developers from manual and tedious work often associated with service deployment. We release the platform as an open-source project on GitHub under Apache 2.0 license, with the aim that it will facilitate and streamline more large-scale ML applications and research projects.

% What are the problem you want to address?

% What kind of approach you are using?

% What are your results?

% What are the impacts of your work?
  
\end{abstract}





\maketitle

\section{Introduction}
\label{sec:introduction}

% context: what is the context of your work

Machine Learning (ML) techniques, especially Deep Learning (DL), have been widely adopted into multimedia applications, ranging from video analysis to artwork generation. To meet the needs of developing ever sophisticated ML applications, both academia and industrial researches have invested a lot of resources and efforts to build systems for speeding up the process of deploying ML as a service (MLaaS). We refer to those systems as ``System for ML" \cite{ratner2019sysml}, which supports the entire ML lifecycle including large-scale distributed training, model deployment, and online inference serving. While most efforts have been focusing on training and serving systems, there is a growing need for efficient, streamlined, and automated model deployment systems.

% challenges: what are the challenges based on the context
% Existing efforts: what are the existing efforts to address those challenges
The development cycle of deploying a MLaaS is often a long and arduous journey. According to \cite{howlongmodel}, 40\% of companies need to spend more than one month deploying a ML model into production while only 14\% can finish this in one week or less. First, developers need to take consideration of a variety of factors such as model running environment and learning parameters to ensure the model performance in the production environment. Second, developers need to very fluent with the architecture of web services such as RESTful or gRPC and write much boilerplate code for different services. Third, to maintain a scalable MLaaS, the developer also need to have a full grasp of infrastructure knowledge of the targeted cloud environment. To address these issues, many systems have been developed, as summarized in Table \ref{tab:system_compare}. A common thread among these systems is the use of containerization techniques with Docker \cite{merkel2014docker} for simplifying the deployment.

\begin{table*}[ht]
\centering
\caption{Comparison of several model deployment frameworks.}
\label{tab:system_compare}
\begin{tabular}{lllllllll}
\hline
Project                                                          & \begin{tabular}[c]{@{}l@{}}Open\\ Source\end{tabular} & \begin{tabular}[c]{@{}l@{}}Model\\ Management\end{tabular} & \begin{tabular}[c]{@{}l@{}}Multi\\ Framework\end{tabular} & Conversion & Profiling  & Dockerization & \begin{tabular}[c]{@{}l@{}}Multi Serving\\ System\end{tabular} & Monitoring \\ \hline
DLHub \cite{chard2019dlhub}                                                           &                                                       & \checkmark                                                 & \checkmark                                                &            &            & \checkmark    & \checkmark                                                     & \checkmark \\ \hline
ModelDB \cite{modeldb}                                                          & \checkmark                                            & \checkmark                                                 & \checkmark                                                &            &            & \checkmark    &                                                                & \checkmark \\ \hline
ModelHub.AI \cite{modelhubai}                                                      & \checkmark                                                       & \checkmark                                                 & \checkmark                                                &            &            & \checkmark    &                                                                &            \\ \hline
Cortex \cite{cortex}                                                           & \checkmark                                            &                                                            & \checkmark                                                &            &            & \checkmark    & \checkmark                                                     & \checkmark \\ \hline
\textbf{MLModelCI}                                                        & \checkmark                                            & \checkmark                                                 & \checkmark                                                & \checkmark & \checkmark & \checkmark    & \checkmark                                                     & \checkmark \\ \hline
\end{tabular}
\end{table*}

% Gaps: what are the shortcomings of existing efforts
Though these systems are widely used, few of them meet the requirements of the real industrial deployment scenario. First, the trained models with python scripts can not be directly deployed to the production environment due to the performance issue. Usually, engineers need to convert a newly trained model to an optimized and hardware-dependent format such as TensorRT \cite{tensorrt} or ONNX \cite{onnx},  to reap the maximal performance benefits of the invested hardware (e.g. graphic processing unit (GPU) and tensor process unit (TPU). Second, the current available systems provide few insights into automatically balancing the trade-off between performance and cost. The performance metrics such as latency and throughput depend on the batch size, the underlying hardware devices, etc. With current solutions, developers still need to go through a manual and time-consuming tuning process.

To address these issues, we develop MLModelCI, a fully open-source platform that provides a one-stop service for optimizing, managing, and deploying MLaaS. The system is informed and motivated by DevOps techniques such as continuous integration (CI) and continuous deployment (CD),  which automates software testing and building before they go online. Specifically, MLModelCI allows users to publish their models and provides management for them. Then it automatically converts models to optimized formats, profiles models under different settings (e.g., batch size and devices), and containerizes models as cloud services.

In MLModelCI, multimedia researchers and engineers have a highly automated solution for robust and efficient MLaaS with a well-designed command line (CLI) toolkit and web interface (i.e., RESTful). Our system reduces the development cycle from weeks or days to hours even minutes. It is written in Python and supported by a lot of industry software frameworks and practice such as Docker, MongoDB, etc., making it easy to be deployed in the cloud and adopted into the existing tool-chains of a team. MLModelCI adheres to best practices in distributed computing and cluster management, by providing a controller to utilize idle workers while maintaining online service quality at the same time.  It also provides much needed features for collaborative ML research, such as bookkeeping models, building demo applications, evaluating model performance, etc.

MLModelCI has been released at \url{https://github.com/cap-ntu/ML-Model-CI} under the license of Apache 2.0. We plan to continuously maintain and upgrade the project to incorporate the rapidly evolving ML techniques. We also build an online discussion community to encourage more researchers and developers to join our effort.

\section{System Highlight}
% \footnote{\url{https://github.com/cap-ntu/ML-Model-CI/tree/master/docs/tutorial}}
MLModelCI provides a complete platform for managing, converting, profiling, and deploying models as cloud services, with well-documented tutorials for all of the related tasks. As such, it serves as a good starting point for researchers to address the gap between models and services and to familiarize themselves with the widely deployed cloud infrastructure. By design, MLModelCI intends to keep the entry barrier as low as possible, so as to make the integration into existing toolchains as efficient and seamless as possible. In this section, we first highlight the key features and advantages of MLModelCI and then compare them with those of other related platforms. 

\subsection{Highlights}

\textbf{Modularity.} Our platform is designed to be as modular as possible from at the get go, allowing a seamless extension to new functions, model formats, and serving systems. We have implemented many features to support model conversion, serving, etc., and provide a lot of examples to show how they can be employed to build efficient MLaaS for various multimedia tasks.

\textbf{Automation.} Upon receiving users' model registration, MLModelCI automates the rest of the tasks. Specifically, it supports a variety of model auto-conversion. It automatically profiles models on available devices. It automates the model binding to existing serving systems and containerization for further deployment.

% As summarized in Table \ref{tab:loc_time}, users only need to write few lines of code (LoC) and spend little time to deploy a MLaaS.

% \begin{table}[!ht]
% \centering
% \caption{Benchmark of ResNet50 MLaaS deployment with MLModelCI and manually using Triton Inference Server.}
% \label{tab:loc_time}
% \begin{tabular}{lll}
% \hline
% Method                                                            & Lines of Code (LoC)                                                                      & \begin{tabular}[c]{@{}l@{}}Time (depend\\ on familiarity)\end{tabular} \\ \hline
% MLModelCI                                                         & \begin{tabular}[c]{@{}l@{}}19 (with conversion,\\ profiling and deployment)\end{tabular} & \begin{tabular}[c]{@{}l@{}}Minitues\\ or hours\end{tabular}            \\ \hline
% \begin{tabular}[c]{@{}l@{}}Triton Inference\\ Server\end{tabular} & \begin{tabular}[c]{@{}l@{}}420 (with no\\ coversion and profiling)\end{tabular}          & \begin{tabular}[c]{@{}l@{}}Days\\ or weeks\end{tabular}                \\ \hline
% \end{tabular}
% \end{table}

\textbf{Elastic.} MLModelCI efficiently uses the available hardware resources by harnessing the idle workers in a serving cluster to complete the profiling. Our platform monitors the hardware status and running models. Once it finds the available hardware resources, it will invoke the profiler to complete the model analysis.

% \textbf{Extensibility.} MLModelCI fully utilizes the Docker software. This allows for rapid extension and development of the system. For instance, if users want to add scaling function for the containerized models as well as the corresponding serving systems, users can employ the Kubernetes without affecting the existing structure and dataflow.

% \textbf{Python based.} For ease-of-use and integrating with existing industry platforms and research models, MLModelCI is built by using Python language. Python is the primary language for researchers to rapidly prototype and train models. Thus it is very natural and efficient to load these models with our platform. Also, many software packages such as Spark have python bindings, making the collaboration seamless.

% \vspace{-1pt}
\subsection{Comparison to Related Platforms}

We provide a survey of related platforms in Table \ref{tab:system_compare}. MLModelCI differs from these platforms in two important aspects:

(1) We focus on supporting online serving rather than recording training experiments. While most of the ModelDB related work focuses on assisting users to log experiments for hyper-parameter searching and can be integrated into training systems, our system focuses on the scenario where the models have been trained successfully and waited to be deployed for online serving. 

(2) Our system not only containerizes models as services but also optimizes and tests them before they go online. By automatically converting and profiling models, MLModelCI ensures a robust and efficient MLaaS and thus saves users' effort and resources.

% Designed as a bridge between training systems and serving systems, it automates the model binding to state-of-the-art serving systems and monitors the online service, making the MLaaS more stable.  

\section{System Architecture}

\begin{figure}[!ht]
  \centering
  \includegraphics[width=0.9\linewidth]{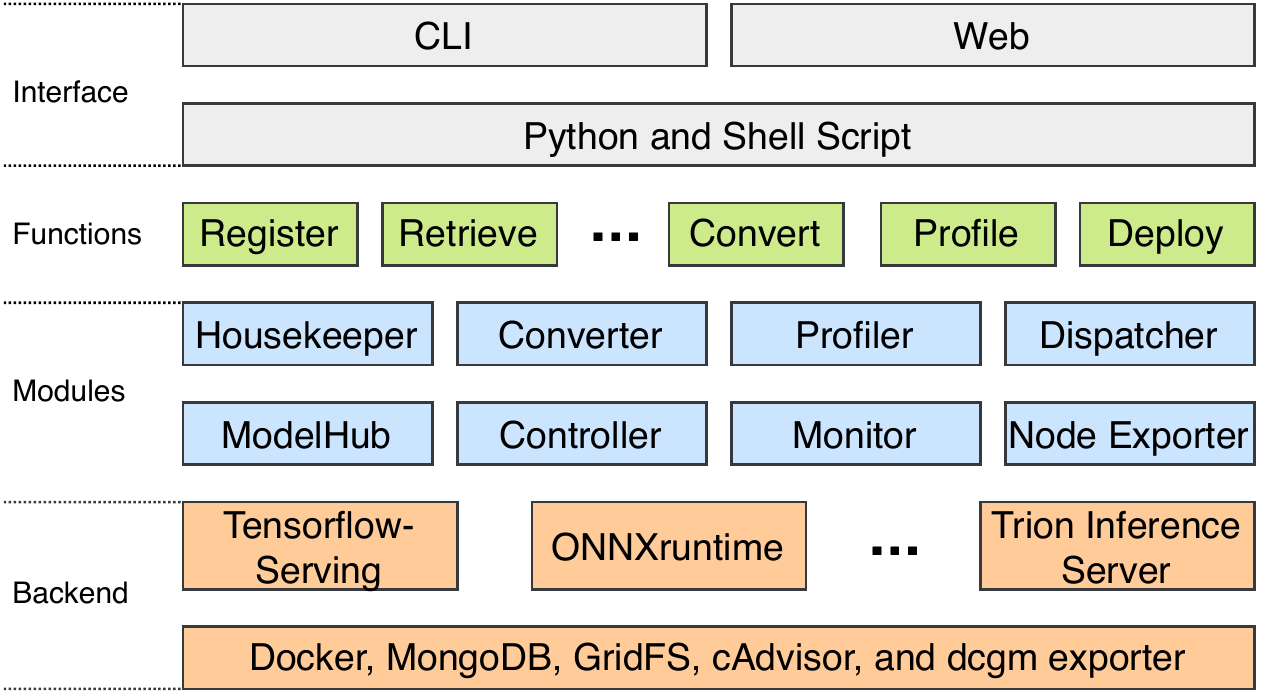}
  \caption{MLModelCI architecture.}
  \label{fig:sys_arch}
  \Description{system architecture}
\end{figure}

In this section, we present the MLModelCI architecture, as illustrated in Figure \ref{fig:sys_arch}. We first summarize the system workflow and then give a detailed description of the functionalities of the core modules.

\begin{figure}[t]
  \centering
  \includegraphics[width=\linewidth]{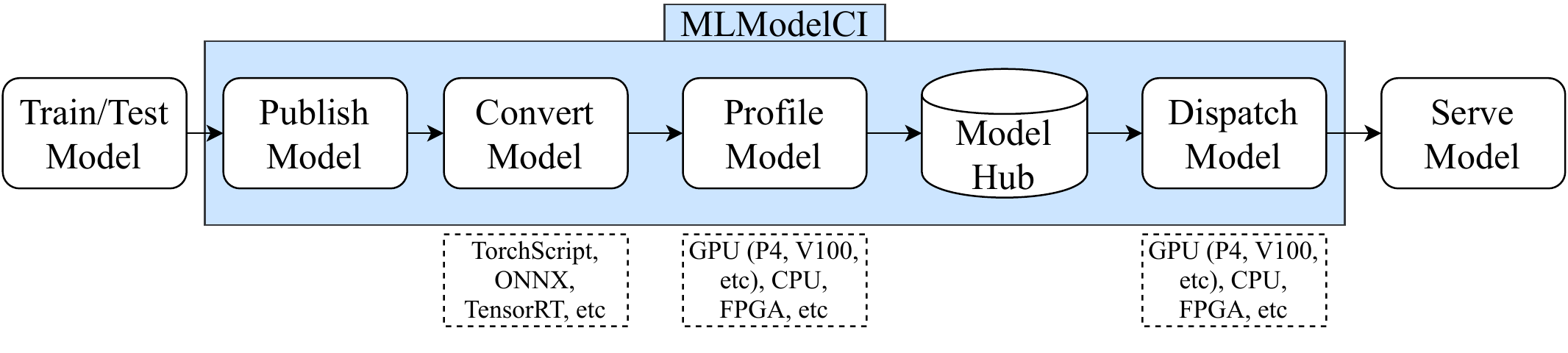}
  \caption{Workflow for a MLaaS deployment task}
  \label{fig:ml_lifecycle}
  \Description{modelci_workflow}
\end{figure}

\textbf{Workflow.} Figure \ref{fig:ml_lifecycle} shows a typical workflow of building a MLaaS with MLModelCI. First, a model weight file associated with a configuration file that contains model registration information (e.g. accuracy, architecture, task, etc.), is published to our system via a simple \texttt{register} API. Next, the system automatically invokes the conversion function to optimize the model, and the profiling function to evaluate model performance on different devices. The information is recorded to guide users to choose the proper batch size, devices, etc., during deployment. At the same time, the model is bound to a serving system and containerized as a service. Finally, users can dispatch the service to a specific device with the help of the \texttt{deploy} API.

Controller, accepting both hardware and running model status, is designed to manage the whole workflow and utilize the idle resources. Non-experts can rely on its automation directly; while experts can customize it for more fine-grained control.

\subsection{ModelHub}

MLModelCI stores models in the modelhub where a model is abstracted into three parts - basic information, dynamic profiling information and a model weight file. Specifically, the basic information includes the model name, the training dataset, accuracy, etc. The dynamic profiling information refers to the runtime performance (e.g., throughput and latency) which is coupled with many aspects (e.g. devices) and is acquired during the profiling process. In practice, this information is more important for guiding online deployment since many papers have mentioned that static information such as FLOPs is often inconsistent with the real speed. 

To persist the information related to a trained model, we adopt MongoDB, a document-based and ease-to-use database, as the storage backend. Meanwhile, its built-in GridFS, a distributed file service, supports large-capacity storage, which is very useful for storing large model weight files. Since the data structure is highly abstracted, users can choose their existing database schema and architecture and applied it to our system with ease. 
% Since the data structure is highly abstracted, users can choose their existing database schema and architecture and applied it to our system with ease.

\subsection{Housekeeper.}

The housekeeper of MLModelCI is the essence of the model management - a team may produce hundreds of models a day. The housekeeper has four key responsibilities for the management and they are encapsulated into four APIs. (1) \texttt{register} accepts a \textit{YMAL} file contains model basic information and a model file from users. Two parameters, conversion, and profiling, can be set to trigger automation processing. (2) \texttt{retrieve} takes the inputs and search for related models to list their information. (3) \texttt{update} is for revising the stored model information. (4) \texttt{delete} is to delete a model from our modelhub.

% \begin{table}[ht]
% \caption{MLModelCI APIs.}
% \begin{tabular}{ll}
% \hline
% API      & Function Description                         \\ \hline
% \texttt{register} & upload model weight file and publish a model \\ 
% \texttt{retrieve} & search models and list their information     \\ 
% \texttt{update}   & update a published model information         \\ 
% \texttt{delete}   & delete a model                               \\ \hline
% \texttt{convert}  & convert a model into optimized format        \\ 
% \texttt{profile}  & profile a model on different devices         \\ \hline
% \texttt{deploy}   & deploy a model to a device                   \\ \hline
% \end{tabular}
% \label{tab:api}
% \end{table}

\subsection{Converter}
\label{sec:converter}

The module focuses on automatically converting research models to serialized and optimized models from Python code rather than carrying out conversion between frameworks (as discussed in MMdnn \cite{MMdnn}). The output models from our converter are independent of the programming language and can be deployed in the production environment.
% \begin{table}[!ht]
% \caption{Supported model conversion formats and the corresponding serving systems.}
% \label{tab:framework-converted-serving}
% \begin{tabular}{lll}
% \hline
% \textbf{Framework}          & \textbf{Converted Format} & \textbf{Serving System}    \\ \hline
% \multirow{2}{*}{PyTorch}    & ONNX                      & ONNX  runtime              \\  
%                             & TorchScript               & Self-defined container     \\
% \multirow{2}{*}{TensorFlow} & Tensorflow SavedModel     & TensorFlow-Serving         \\  
%                             & TensorRT                  & Trion Inference Server \\ \hline
% \end{tabular}
% \end{table}

The whole pipeline is as follows. First, the converter obtains the framework information (e.g., PyTorch) of a registered model. Then according to the information, it translates the model including its structure and pretrained weight to a deployable protobuf file by utilizing the corresponding toolkit that has been incorporated into our system. For instance, a PyTorch model can be converted to TorchScipt and ONNX formats; while a Tensorflow can be converted to SavedModel and TensorRT formats. These models can be bound to serving systems for further optimized during runtime.
%  The output models from our converter are independent of the programming language and can be deployed in the production environment. 

\subsection{Profiler}
\label{sec:profiler}

MLModelCI runs models on the underlying devices in the clusters, and collects, aggregates, and processes running model performance. Specifically, there are six indicators that will be acquired, peak throughput, P50-latency, P95-latency, P99-latency, GPU memory usage, and GPU computation utilization. 

To get real model performance in practice, the profiler simulates the real service behavior by invoking a gRPC client and a model service. In particular, the profiler contains many build-in clients and upon it receives a profiling signal, it starts the corresponding client and invokes the dispatcher to deploy a MLaaS. It then sends test data from the client to the service with a variety of batch sizes and serving systems on different devices. Users can have hundreds of combinations available, which is very useful for setting parameters for online services.

\begin{figure*}[t]
  \centering
  \includegraphics[width=0.9\linewidth]{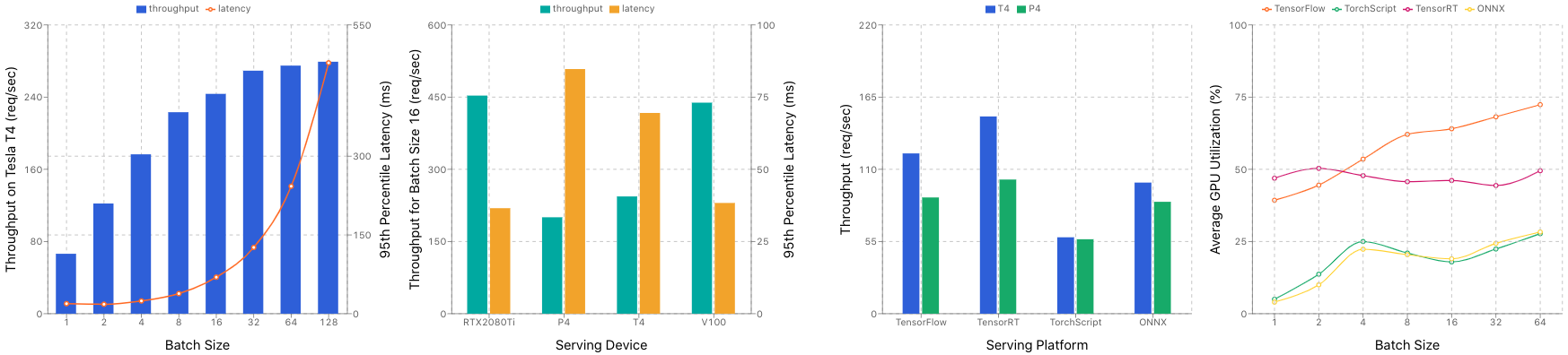}
  \caption{Highlighted profiling results. Try more demos yourself online}
  \label{fig:resnet50_profiling}
  \Description{model profiling}
\end{figure*}

\subsection{Dispatcher}
The dispatcher launches a serving system (e.g. Tensorflow-Serving) to load a model in a containerized manner and dispatches the MLaaS to a device. We have dockerized many widely used serving systems as shown in Figure \ref{fig:sys_arch} to support different model formats. MLModelCI supports building two kinds of web service, RESTful and gRPC. The former, built on top of the HTTP, is very useful for applications that only need one model. In comparison, gRPC is designed for low latency and high throughput communication, and supports to build a service with multiple models well.

\subsection{Monitor \& Node Exporter}

Our monitor collects and aggregates running model container performance. We use cAdvisor as the backend and get container status from it periodically. The information includes resource usage (e.g., GPU memory, CPU usage, etc.), and network statistics. 

The node exporter collects hardware status and exposes them to our system. The node exporter is based on two software, prometheus and dcgm exporter. The former can aggregate the CPU and network utilization, and the later can collect the GPU metrics.
\subsection{Controller}

The controller receives data from the monitor and node exporter, and controls the whole workflow of our system. First, it guides the profiler to evaluate models when devices are idle periodically. Users choose the threshold of device utilization that constitutes a system being considered as idle. For instance, users can set this threshold as 40\%. If the utilization of a GPU is higher than this threshold, the model can not be profiled on it now. Second, it helps to automatically set up a MLaaS to available devices.

\section{Demonstration}

% \begin{figure}[t]
%   \centering
%   \includegraphics[width=1.0\linewidth]{exp/front_end.png}
%   \caption{The housekeeper fontend.}
%   \label{fig:mangement_frontend}
%   \Description{Model Management}
% \end{figure}

This section illustrates the MLaaS deployment supported by MLModelCI. We use widely deployed models for multimedia analysis as examples and relevant source code is released on GitHub\footnote{\url{https://github.com/cap-ntu/mlmodelci_mm_demo}}.

\begin{figure}[!ht]
\centering
\begin{subfigure}{0.24\textwidth}
\centering
\includegraphics[width=1.0\linewidth]{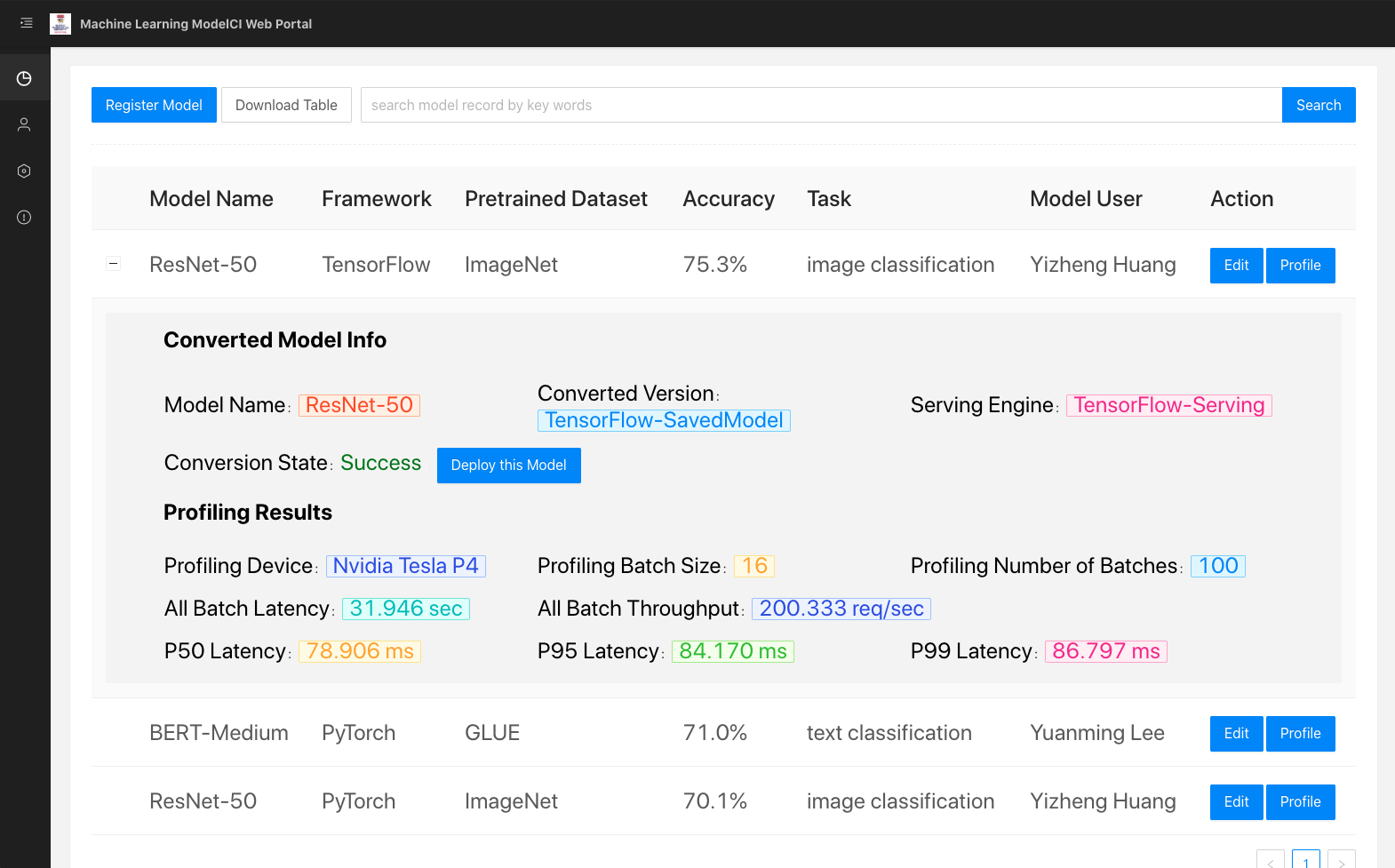}
\caption{Housekeeper Frontend}
\label{fig:mangement_frontend}
\end{subfigure}%
\begin{subfigure}{0.255\textwidth}
\centering
\includegraphics[width=1.0\linewidth]{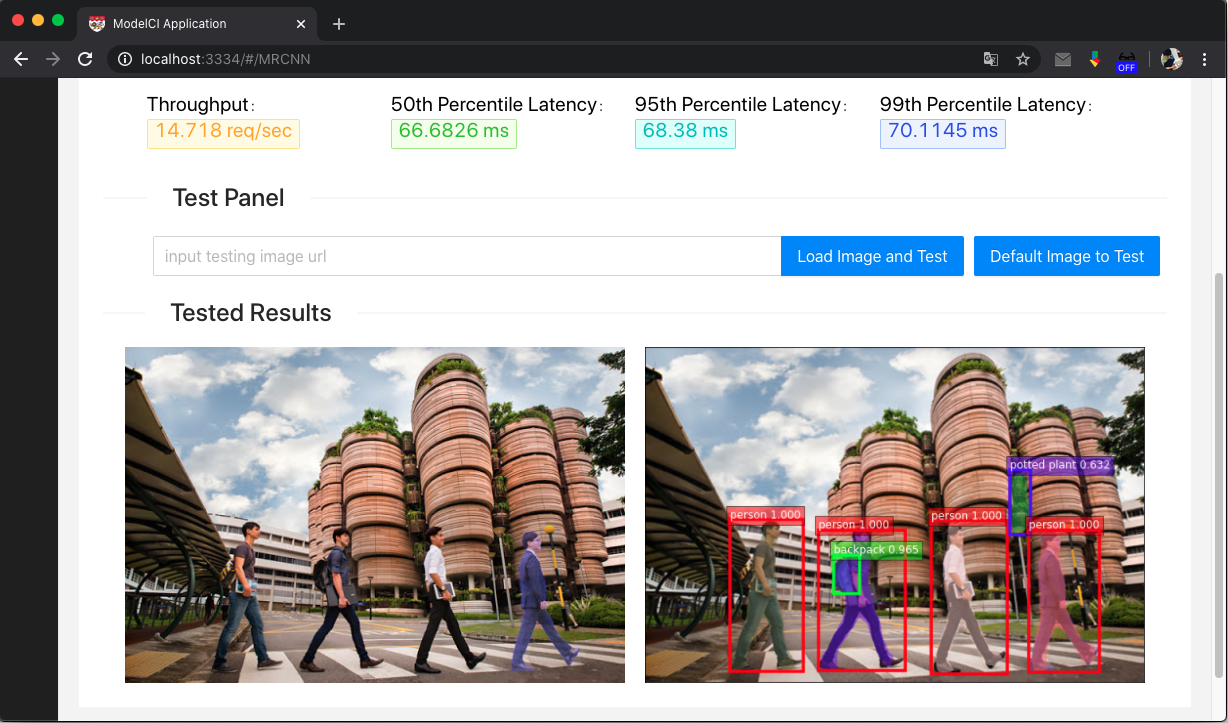}
\caption{Mask R-CNN Service}
\label{fig:maskrcnn}
\end{subfigure}

\caption{Web interface and a demo application.}
\label{fig:two_apps}
\end{figure}

\subsection{Model Publishing \& Conversion}
% MLModelCI has an online tutorial\footnote{\url{https://github.com/cap-ntu/ML-Model-CI/tree/master/docs/tutorial}} to guide users to build a MLaaS step by step.
To deploy a MLaaS, the first step is to publish a model to our system. We use a widely deployed image classification model - ResNet50 - as an example. A registration \textit{YAML} file associated with a model file is prepared. After uploading them to MLModelCI, users can manage the model as shown in Figure \ref{fig:mangement_frontend}. More detailed evaluation results are presented after the conversion and profiling. As shown in Figure \ref{fig:mangement_frontend}, there is a successful converted model and a TensorflowSaved model served by Tensorflow-Serving. The model runtime performance is also presented, indicating that profiling has finished.

\subsection{Service Profiling}

We now use a real-world example to illustrate the necessity of the model (service) profiling\footnote{\url{https://cap-ntu.github.io/mlmodelci_mm_demo/}}. MLModelCI manages to obtain as much as possible information and generates a comparison report. As shown in Figure \ref{fig:resnet50_profiling}, the model runtime performance is determined by many as aspects (from left to right, batch size, devices and serving platforms), and their resource usage such as the GPU utilization varies. MLModelCI provides all of the information to users to help build a more cost-effective solution.

% (1) As the batch size increases, the throughput first increases then becomes stable while the P95 latency increases steadily. For those real-time services, choosing an appropriate batch size is vital. (2) When the batch size is fixed, the choice of hardware devices affects the performance a lot. Since hardware prices vary, it is better to select a cheaper device that meets the latency requirement. (3) The performance is also influenced by the model formats with the corresponding serving systems. As we already converted models for users, they can easily pick up a more efficient solution. (4) GPU utilization is also presented. If the GPU utilization of service keeps low, users may decide to share the device. 

% \begin{figure}[!ht]
%   \centering
%   \includegraphics[width=0.8\linewidth]{exp/cost.png}
%   \caption{Cost estimation of a variety of devices. Users can define the prices by themselves and MLModelCI calculated the total cost.}
%   \label{fig:cost_estimation}
%   \Description{Model Management}
% \end{figure}

\subsection{MLaaS Deployment}

We deploy another widely used model in the multimedia application: Mask R-CNN \cite{he2017mask},  as shown in Figure \ref{fig:maskrcnn}. By following the official deployment code from TensorFlow-Serving, developers need to write more than 500 lines of code (LoC) to complete the Mask R-CNN MLaaS building, not including any conversion and profiling. The whole deployment works may take days or weeks depending on developers' familiarity with TensorFlow-Serving's deployment. In contrast, with the help of MLModelCI, users only need to write about 20 LoC to complete the deployment.

% The first is BERT \cite{devlin2018bert} model. It takes text as inputs and represents data for downstream applications. We build a sentiment analysis service based on it as shown in Figure \ref{fig:bert}. The second is Mask R-CNN \cite{he2017mask}. The model is developed to do both object detection and segmentation. We build a Mask R-CNN service as shown in Figure \ref{fig:maskrcnn}. MLModelCI already makes full support for these widely deployed models which covers a lot of scenarios. Also, the services are dockerized so they can be scaled up easily according to the online workload with the help of Kubernetes.

\section{Conclusion}

In this paper, we described MLModelCI, an automated and easy-to-use platform for building efficient and robust MLaaS in cloud. It is designed to free researchers and developer from the tedious and error-prone model deployment works. We develop and automate many much needed functions/features, thus bridging the gap between existing training and serving systems. We demonstrate the usability of the system with representative case studies. We plan to continually maintain the system for further improving both the speed and the efficiency of ML model delivery and deployment.

\bibliographystyle{ACM-Reference-Format}
\bibliography{sample-base}

% \appendix









\end{document}